\documentclass[aps,prl,showpacs,twocolumn,superscriptaddress]{revtex4}
\usepackage{bm,color}
\usepackage{graphicx}
\usepackage{amsmath}

\begin{document}
\title {Magnetoelectric Resonances and Predicted Microwave Diode Effect of Skyrmion Crystal in Multiferroic Chiral-Lattice Magnet}

\author{Masahito Mochizuki}
\email{mochizuki@ap.t.u-tokyo.ac.jp}
\affiliation{Department of Applied Physics, The University of Tokyo,
Tokyo 113-8656, Japan}
\affiliation{Institute of Theoretical Physics, University of Cologne,
D-50937 Cologne, Germany}

\author{Shinichiro Seki}
\affiliation{Department of Applied Physics and Quantum Phase Electronics 
Center, The University of Tokyo, Tokyo 113-8656, Japan}


\begin{abstract}
We theoretically discover that unique eigenmodes of skyrmion crystal (SkX) are not only magnetically active to ac magnetic field ($\bm H^\omega$) but also electrically active to ac electric field ($\bm E^\omega$) in a multiferroic chiral-lattice magnet Cu$_2$OSeO$_3$, which amplifies the dynamical magnetoelectric coupling between $\bm E^\omega$ and the spin texture. The resulting intense interference between their electric and magnetic activation processes can lead to unprecedentedly large diode effect on the microwave, i.e., its absorption by SkX changes up to $\sim$20$\%$ when the incident direction is reversed. Our results demonstrate that the skyrmion can be a promising building block for microwave devices.
\end{abstract}
\pacs{76.50.+g, 75.70.Ak, 75.10.Hk, 75.78.-n}
\maketitle
Skyrmion, a topological vortex-like swirling spin texture~\cite{Skyrme62}, is now attracting a great deal of interest. It was predicted that the skyrmion and its crystallized form, so-called skyrmion crystal (SkX), are realized in chiral-lattice magnets without inversion symmetry through competition between ferromagnetic (FM) and Dzyaloshinskii-Moriya (DM) interactions under magnetic field $\bm H$~\cite{Bogdanov89,Rossler06}. Quite recently, the SkX phase was indeed observed in metallic B20 alloys such as MnSi~\cite{Muhlbauer09,Pappas09,Pfleiderer10,Adams11}, Fe$_{1-x}$Co$_x$Si~\cite{Munzer10,YuXZ10N} and FeGe~\cite{YuXZ10M} as well as in an insulating magnet Cu$_2$OSeO$_3$~\cite{Seki12a,Adams12,Seki12b} by small-angle neutron-scattering experiments (SANS) and Lorentz transmission electron microscopy (LTEM).

Since then, several experiments have been performed, and have reported intriguing transport properties in SkX~\cite{Binz08,Neubauer09,Kanazawa11,Pfleiderer10b,Schulz12,Huang12} and electric control of skyrmions with spin-polarized current~\cite{Jonietz10,YuXZ12} in the {\it metallic} systems. The emergence of spin-driven ferroelectric polarization $\bm P$ has been observed in the {\it insulating} SkX phase of Cu$_2$OSeO$_3$~\cite{Seki12a,Belesi12,Seki12c}, and the electric-field control of this {\it multiferroic} skyrmion texture was experimentally demonstrated~\cite{WhiteCD}. There the research interest, more or less, comes from the possible application to next-generation spintronics devices. However a lot of researchers are presaging further potentiality in the skyrmions. Nevertheless the sorts of experimental works are quite limited, i.e., observations by means of LTEM or SANS and transport measurements only.

In this Letter, we theoretically propose a brand new direction to the research on skyrmions from the viewpoints of microwave functionalities and dynamical phenomena at GHz frequencies. We discover that collective rotational and breathing motions of skyrmions in SkX can be resonantly activated not only by ac magnetic field ($\bm H^\omega$) but also by ac electric field ($\bm E^\omega$) as unique eigenmodes of SkX in multiferroic chiral-lattice magnets. These resonances amplify the dynamical coupling of underlying spin texture with $\bm E^\omega$ and $\bm H^\omega$, and the resulting intense interference between the electric and magnetic activation processes can lead to unprecedentedly large directional dichroism of electromagnetic (EM) wave in Cu$_2$OSeO$_3$, i.e., its absorption by SkX changes up to $\sim$20$\%$ depending on the sign of its incident direction. This effect is enhanced especially at eigenfrequencies of the aforementioned skyrmion resonances ($\sim$GHz), and can work as an efficient microwave diode. Currently, most of the microwave-device functions are achieved using designed combinations of waveguides, circuits and elements made of ferrites with ferrimagnetic order~\cite{Gurevich96}. Our finding provides a guideline for designing new microwave devices such as magnetically tunable isolator. Our work will be a trigger for a broad-based quest for novel functions of skyrmions and related spin textures.

\begin{figure}
\includegraphics[scale=1.0]{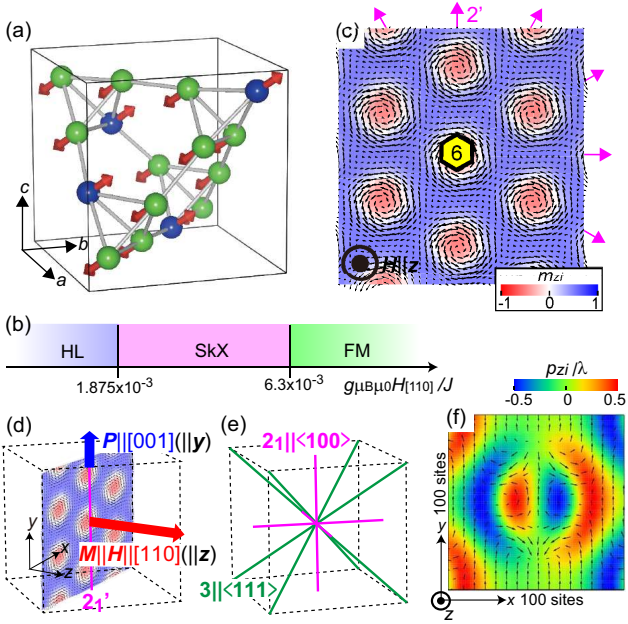}
\caption{(color). (a) Spin structure of Cu$_2$OSeO$_3$, composed of tetrahedra of four Cu$^{2+}$-ions ($S$=1/2) with three-up and one-down spins. (b) Phase diagram of the spin model~(\ref{eqn:model}). Here HL, SkX, and FM denote helical, skyrmion-crystal, and ferromagnetic phases, respectively. (c) Spin structure in the SkX phase, which possesses a six-fold rotation axis, 6, and six two-fold rotation axes followed by time reversal, $2'$. Arrows represent the in-plane spin components. (d) Under $\bm H$$\parallel$[110], the system becomes polar along [001] and the emergence of ferroelectric polarization $\bm P$$\parallel$[001] is allowed. (e) Symmetry axes in the Cu$_2$OSeO$_3$ crystal, which belongs to the P2$_1$3 space group: three-fold rotation axes, 3, along $\left< 111 \right>$ and two-fold screw axes, $2_1$, along $\left< 100 \right>$. (f) Real-space configurations of local electric polarizations $\bm p_i$ in the skyrmion under $\bm H$$\parallel$[110].}
\label{Fig1}
\end{figure}
The magnetic structure of Cu$_2$OSeO$_3$ is composed of tetrahedra of four Cu$^{2+}$ ($S$=1/2) ions as shown in Fig.~\ref{Fig1}(a). Recent powder neutron diffraction~\cite{Bos08} and NMR~\cite{Belesi10} experiments suggested that three-up and one-down (uuud) type collinear spin arrangement is realized on each tetrahedron below $T_{\rm c}$$\sim$58 K. We regard this four-spin assembly as a magnetic unit, and treat it as a classical vector spin $\bm m_i$ whose norm $m$ is unity. We employ a classical Heisenberg model on a cubic lattice~\cite{Bak80,YiSD09,HanJH10} to describe the magnetism in thin specimen of Cu$_2$OSeO$_3$, which contains the FM-exchange interaction, the DM interaction\cite{YangJH12} and the Zeeman coupling to the external $\bm H$ normal to the plane. The Hamiltonian is given by,
\begin{eqnarray}
\mathcal{H}_0&=&
-J \sum_{<i,j>} \bm m_i \cdot \bm m_j
-D \sum_{i,\hat{\bm \gamma}} 
\bm m_i \times \bm m_{i+\hat{\bm \gamma}} \cdot \hat{\bm \gamma} 
\nonumber \\
&-&g\mu_{\rm B}\mu_0 H_z \sum_i m_{iz},
\label{eqn:model}
\end{eqnarray}
where $g$=2, and $\hat{\bm \gamma}$ runs over $\hat{\bm x}$ and $\hat{\bm y}$. We set the ratio $D/J$=0.09, for which the periodicity in the SkX phase becomes $\sim$99 sites. Since the distance between adjacent Cu-ion tetrahedra is $\sim$5 $\AA$, this periodicity corresponds to the skyrmion diameter of $\sim$50 nm in agreement with the LTEM observation~\cite{Seki12a}. All the spin textures considered here are slowly varying, which can be described by a continuum spin model. It justifies our treatment based on a lattice spin model after dividing the space into square meshes and the coarse graining of magnetizations.

We first analyze the model~(\ref{eqn:model}) using the replica-exchange Monte-Carlo technique, and obtain phase diagrams at low temperature ($T$) shown in Fig.~\ref{Fig1}(b). The SkX phase emerges in the range $1.875\times10^{-3}<|g\mu_{\rm B}\mu_0 H_z/J|<6.3\times10^{-3}$, sandwiched by the helical and FM phases in agreement with the experiment for thin-plate samples~\cite{Seki12a}. The skyrmions are crystallized into a triangular lattice and the magnetic moments $\bm m_i$ direct antiparallel (parallel) to $\bm H$ at the center (periphery) of each skyrmion as shown in Fig.~\ref{Fig1}(c).

For the SkX state formed under $\bm H$$\parallel$[110] as shown in Fig.~\ref{Fig1}(d), the emergence of $\bm P$$\parallel$[001] perpendicular to the net magnetization $\bm M$($\parallel$$\bm H$) is expected from symmetry consideration. As shown in Fig.~\ref{Fig1}(e), the crystal structure of Cu$_2$OSeO$_3$, which belongs to a non-polar space group, P2$_1$3, possesses three-fold rotation axes, 3, along $\left< 111 \right>$, and $2_1$-screw axes along $\left< 100 \right>$. The spin texture in the SkX phase is also non-polar with a six-fold rotation axis, 6, along $\bm H$, and two-fold rotation axes followed by time reversal, $2'$, normal to $\bm H$ as shown in Fig.~\ref{Fig1}(c). When the SkX sets in under $\bm H$$\parallel$[110] on the Cu$_2$OSeO$_3$ crystal, most of the symmetries should be broken, and only the $2_1'$-axis ($\parallel$[001]) normal to $\bm H$ survives as shown in Fig.~\ref{Fig1}(d). Consequently, the system becomes polar along [001]. Indeed the emergence of $\bm P$$\parallel$[001] under $\bm H$$\parallel$[110] was experimentally observed~\cite{Seki12c}. Here we define the Cartesian coordinates, $\bm x$$\parallel$[$\bar 1$10], $\bm y$$\parallel$$\bm P$$\parallel$[001], and $\bm z$$\parallel$$\bm M$$\parallel$[110] shown in Fig.~\ref{Fig1}(d), for convenience of the following formulations.

The net magnetization $\bm M$ and the ferroelectric polarization $\bm P$ are given by sums of the local contributions as $\bm M$=$\frac{g\mu_{\rm B}}{NV}\sum_{i=1}^{N} \bm m_i$ and $\bm P$=$\frac{1}{NV}\sum_{i=1}^{N} \bm p_i$, respectively, where the index $i$ runs over the Cu-ion tetrahedra with uuud spin pair, $N$ is the number of the tetrahedra, and $V$(=1.76$\times$10$^{-28}$ m$^3$) is the volume per tetrahedron. Because of the cubic symmetry, the local polarization $\bm p_i$ from the $i$th tetrahedron is given using the spin components $m_{ia}$, $m_{ib}$, and $m_{ic}$ in the P2$_1$3 setting as,
\begin{eqnarray}
\bm p_i=\left(p_{ia}, p_{ib}, p_{ic} \right)
= \lambda \left(m_{ib}m_{ic}, m_{ic}m_{ia}, m_{ia}m_{ib} \right).
\end{eqnarray}
We can easily evaluate the local contributions $\bm p_i$ and $\bm m_i$ from each tetrahedron in the ferrimagnetic phase where all the tetrahedra give uniform contributions. Then the coupling constant $\lambda$ is evaluated as $\lambda$=$5.64\times10^{-27}$ $\mu$Cm from the experimentally measured $P_{[001]}$=16 $\mu$C/m$^2$ in the ferrimagnetic phase under $\bm H$$\parallel$[111] at 5 K~\cite{Seki12a}.

Because of this strong coupling between magnetism and electricity, collective oscillations of this SkX can be activated not only magnetically by ac magnetic field $\bm H^\omega$ but also electrically by ac electric field $\bm E^\omega$. As demonstrated below, with the special configuration of $\bm P$$\perp$$\bm M$, both the $\bm H^\omega$ and $\bm E^\omega$ components of EM wave propagating along $\bm P$$\times$$\bm M$ can activate common oscillation modes. To see this, we calculate dynamical magnetic and dielectric susceptibilities,
\begin{eqnarray}
\chi^{\rm mm}_{\alpha \beta}(\omega) =
\frac{M_{\alpha}^{\omega}}{\mu_0 H_{\beta}^{\omega}},
\;\;\;
\chi^{\rm ee}_{\alpha \beta}(\omega) =
\frac{P_{\alpha}^{\omega}}{\epsilon_0 E_{\beta}^{\omega}},
\end{eqnarray}
by numerically solving the Landau-Lifshitz-Gilbert equation using the fourth-order Runge-Kutta method. The equation is given by
\begin{equation}
\frac{d\bm m_i}{dt}=-\bm m_i \times \bm H^{\rm eff}_i 
+\frac{\alpha_{\rm G}}{m} \bm m_i \times \frac{d\bm m_i}{dt},
\label{eq:LLGEQ}
\end{equation} 
where $\alpha_{\rm G}$(=0.04) is the Gilbert-damping coefficient. The effective field $\bm H_i^{\rm eff}$ is calculated from the Hamiltonian $\mathcal{H}$=$\mathcal{H}_0$+$\mathcal{H}'(t)$ as
$\bm H^{\rm eff}_i = -\partial \mathcal{H} / \partial \bm m_i$.
Here the first term $\mathcal{H}_0$ is the model Hamiltonian~(\ref{eqn:model}), while the perturbation term $\mathcal{H}'(t)$ represents a short rectangular pulse of magnetic field or electric field.
After applying the pulse at $t$=0, we calculate $\bm M(t)$ and $\bm P(t)$, and obtain their Fourier transforms $M_{\alpha}^{\omega}$ and $P_{\alpha}^{\omega}$. The calculations are performed using a system of $N$=288$\times$288 sites with the periodic boundary condition.

\begin{figure}
\includegraphics[scale=0.9]{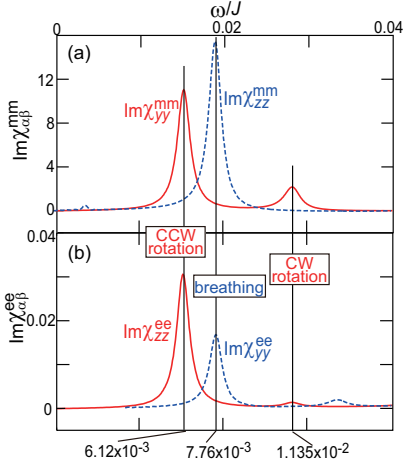}
\caption{(color online). (a) Imaginary parts of the dynamical magnetic susceptibilities, Im$\chi^{\rm mm}_{yy}$ and Im$\chi^{\rm mm}_{zz}$, as functions of frequency $\omega$ at $g\mu_{\rm B}\mu_0H_z/J$=3.75$\times$10$^{-3}$. In Ref.~\cite{Mochizuki12}, the strong (weak) resonance in Im$\chi^{\rm mm}_{yy}$ was ascribed to the counterclockwise (clockwise) rotation mode, while the resonance in Im$\chi^{\rm mm}_{zz}$ was to the breathing mode. (b) Imaginary parts of the dynamical dielectric susceptibilities, Im$\chi^{\rm ee}_{zz}$ and Im$\chi^{\rm ee}_{yy}$, as functions of $\omega$. Above mentioned three magnetically active resonances can be seen in the spectra of dielectric susceptibilities at the same frequencies, indicating their simultaneous electric activities.}
\label{Fig2}
\end{figure}
In Fig~\ref{Fig2}(a), we display imaginary parts of the calculated dynamical magnetic susceptibilities, Im$\chi^{\rm mm}_{yy}(\omega)$ and Im$\chi^{\rm mm}_{zz}(\omega)$, for $\bm H^\omega$$\parallel$$\bm y$ and $\bm H^\omega$$\parallel$$\bm z$. We also plot imaginary parts of the calculated dielectric susceptibilities, Im$\chi^{\rm ee}_{zz}(\omega)$ and Im$\chi^{\rm ee}_{yy}(\omega)$, for $\bm E^\omega$$\parallel$$\bm z$ and $\bm E^\omega$$\parallel$$\bm y$ in Fig~\ref{Fig2}(b). In Im$\chi^{\rm mm}_{yy}$, we find a strong resonance active to $\bm H^\omega$$\parallel$$\bm y$ at $\omega_{\rm R}/J$=6.12$\times$10$^{-3}$, which was ascribed to the counterclockwise rotation mode where all the skyrmion cores in the SkX uniformly rotate in the counterclockwise fashion~\cite{Mochizuki12,Petrova11}. This rotation mode can be seen also in the spectrum of Im$\chi^{\rm ee}_{zz}$ as a peak at the same frequency indicating its simultaneous electric activity to $\bm E^\omega$$\parallel$$\bm z$. The spectrum of Im$\chi^{\rm mm}_{yy}$ has one more resonance at higher frequency $\omega_{\rm R}$=$1.135 \times 10^{-2}J$, which was ascribed to another rotation mode with opposite rotational sense, i.e., the clockwise rotation mode~\cite{Mochizuki12,Petrova11}. We can see a very tiny peak in Im$\chi^{\rm ee}_{zz}$ at the corresponding frequency, indicating its weak electric activity. On the other hand, the spectrum of Im$\chi^{\rm mm}_{zz}$ has a single resonance active to $\bm H^\omega$$\parallel$$\bm z$ at $\omega_{\rm R}/J$=7.76$\times$10$^{-3}$, which was ascribed to the breathing mode where areas of all the skyrmions in the SkX oscillatory expand and shrink in a uniform way~\cite{Mochizuki12,Petrova11}. Again this mode is simultaneously active to $\bm E^\omega$$\parallel$$\bm y$, and the corresponding peak can be seen in Im$\chi^{\rm ee}_{yy}$ at the same frequency. Recent microwave experiment found clear absorptions at these spin-wave resonances, while absorptions at off-resonant frequencies turn out negligibly small~\cite{Onose12}.

\begin{figure}
\includegraphics[scale=1.0]{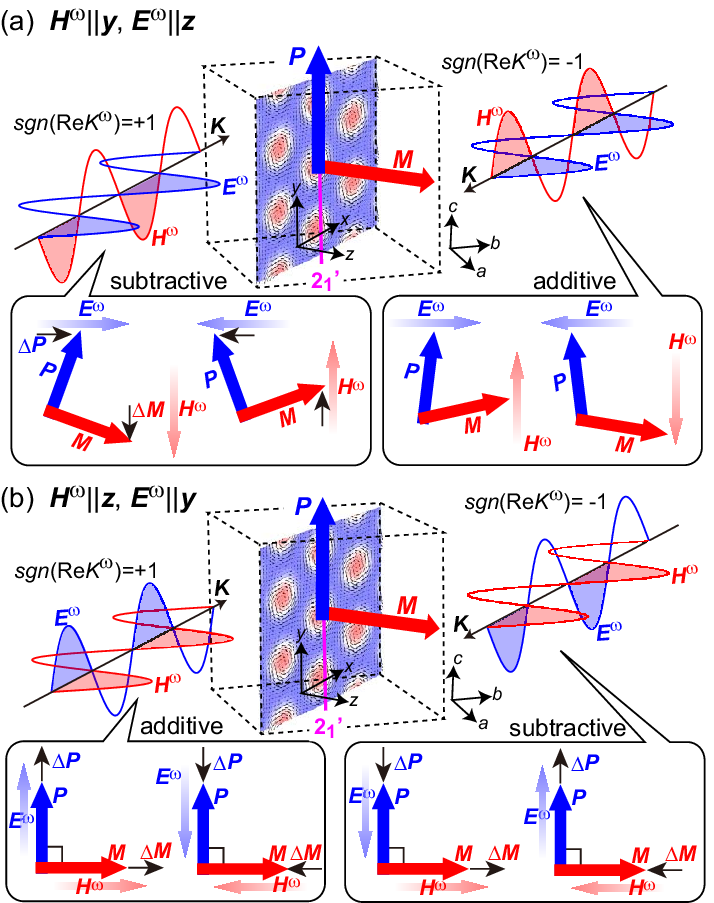}
\caption{(color online). Configurations of the microwave $\bm H^\omega$ and $\bm E^\omega$ components, for which the nonreciprocal directional dichroism is expected when $\bm P$$\parallel$$\bm y$ and $\bm M$$\parallel$$\bm z$ with $\bm P$$\perp$$\bm M$, and $\bm K^\omega$$\parallel$$\bm P$$\times$$\bm M$($\parallel$$\bm x$);
(a)
$\bm E^\omega$$\parallel$$\bm z$ and 
$\bm H^\omega$$\parallel$$\bm y$, and
(b)
$\bm E^\omega$$\parallel$$\bm y$ and
$\bm H^\omega$$\parallel$$\bm z$. 
Cartesian coordinates $\bm x$$\parallel$[$\bar{1}$10], $\bm y$$\parallel$[001], and $\bm z$$\parallel$[110] are defined as shown in the figures where the $z$ axis is parallel to $\bm H$.}
\label{Fig3}
\end{figure}
The presence of collective modes active to both $\bm E^\omega$ and $\bm H^\omega$ is nothing but the source of interesting microwave activity. From Maxwell's equations, we can derive the relation $\bm H^{\omega}$$\parallel$$\bm K^\omega$$\times$$\bm E^{\omega}$ for the EM wave. This relation indicates that relative directions of $\bm H^{\omega}$ and $\bm E^{\omega}$ are determined by the propagation vector $\bm K^\omega$, and their relationship should be reversed upon the sign reversal of $\bm K^\omega$. When the lineally polarized EM wave with $\bm E^\omega$$\parallel$$\bm z$ and $\bm H^\omega$$\parallel$$\bm y$ propagates parallel (antiparallel) to $(\bm P \times \bm M)$$\parallel$$\bm x$ as shown in Fig.~\ref{Fig3}(a), where sgn[Re$K^\omega$]=$+1$ (sgn[Re$K^\omega$]=$-1$) with $\bm K^\omega$=$K^\omega \hat{\bm x}$, the oscillation of $\bm P$ induced by $\bm E^\omega$ and that of $\bm M$ by $\bm H^\omega$ contributes in a subtractive (an additive) way to the collective oscillation, which results in weaker (stronger) absorption of the EM wave. Such a nonreciprocal absorption of EM-wave is expected also for $\bm E^\omega$$\parallel$$\bm y$ and $\bm H^\omega$$\parallel$$\bm z$ as shown in Fig.~\ref{Fig3}(b).

\begin{figure}[t]
\includegraphics[scale=1.0]{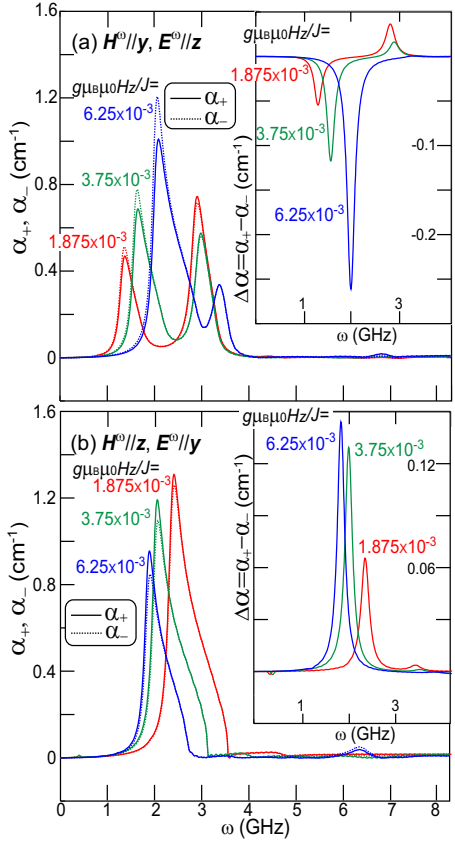}
\caption{(color online).  Calculated absorption coefficients, $\alpha_{+}(\omega)$ and $\alpha_{-}(\omega)$ for microwaves with sgn(Re$K^\omega$)=$+1$ and sgn(Re$K^\omega$)=$-1$, respectively, at several values of $H_z$ in the cases of (a) $\bm H^{\omega}$$\parallel$$\bm y$ and $\bm E^{\omega}$$\parallel$$\bm z$ and (b) $\bm H^{\omega}$$\parallel$$\bm z$ and $\bm E^{\omega}$$\parallel$$\bm y$.}
\label{Fig4}
\end{figure}
To study the microwave absorption and NDD quantitatively, we start with the following Fourier-formed Maxwell's equations for materials with $\bm M$ and $\bm P$~\cite{Kezsmarki11},
\begin{eqnarray}
 \omega {\bm B}^\omega = {\bm K^\omega} \times {\bm E}^\omega,
\;\;\;\;
-\omega {\bm D}^\omega = {\bm K^\omega} \times {\bm H}^\omega
\label{eq:FTMaxwellEq}
\end{eqnarray}
where
\begin{eqnarray}
{\bm B}^{\omega} &=& \mu_0
(\hat{\mu}^\infty {\bm H}^{\omega}+{\bm M}^{\omega}),
\;\;
{\bm D}^{\omega} = \epsilon_0 \hat{\epsilon}^\infty 
{\bm E}^{\omega}+{\bm P}^{\omega},
\label{eq:B&D}
\\
{\bm M}^{\omega} &=& \mu_0 \hat{\chi}^{\rm mm}(\omega) {\bm H}^{\omega}
+ \hat{\chi}^{\rm me}(\omega) 
\sqrt{\frac{\epsilon_0}{\mu_0}} {\bm E}^{\omega},
\\
{\bm P}^{\omega} &=& \epsilon_0 \hat{\chi}^{\rm ee}(\omega) {\bm E}^{\omega}
+ \hat{\chi}^{\rm em}(\omega)
\sqrt{\epsilon_0 \mu_0} {\bm H}^{\omega}.
\label{eq:P&M}
\end{eqnarray}
As discussed above, the NDD of microwaves with $\bm K^\omega$=$K^\omega \hat{\bm x}$ is expected for the following configurations of $\bm H^\omega$ and $\bm E^\omega$; (i) $\bm H^\omega$$\parallel$$\bm y$ and $\bm E^\omega$$\parallel$$\bm z$, and (ii) $\bm H^\omega$$\parallel$$\bm z$ and $\bm E^\omega$$\parallel$$\bm y$. We introduce the complex refractive index $N(\omega)$=$n(\omega)+i\kappa(\omega)$ which is related with $K^\omega$ as $K^\omega$=$\frac{\omega}{c} N(\omega)$. We solve Eqs.~(\ref{eq:FTMaxwellEq}) and obtain,
\begin{eqnarray}
N(\omega) &\sim&  \sqrt{
[{\epsilon_{zz}}^\infty + \chi^{\rm ee}_{zz} (\omega)]
[{\mu_{yy}}^\infty + \chi^{\rm mm}_{yy}  (\omega)]}
\nonumber \\
&-& sgn({\rm Re}K^\omega) [\chi^{{\rm me}}_{yz} (\omega)
+ \chi^{{\rm em}}_{zy} (\omega)]/2,
\label{eqn:CRI1}
\end{eqnarray}
and 
\begin{eqnarray}
N(\omega) &\sim&  \sqrt{
[{\epsilon_{yy}}^\infty + \chi^{\rm ee}_{yy} (\omega)]
[{\mu_{zz}}^\infty + \chi^{\rm mm}_{zz}  (\omega)]}
\nonumber \\
&+& sgn({\rm Re}K^\omega) [\chi^{{\rm me}}_{zy} (\omega)
+ \chi^{{\rm em}}_{yz} (\omega)]/2,
\label{eqn:CRI2}
\end{eqnarray}
for Cases (i) and (ii), respectively.
These expressions contain the sign of Re$K^\omega$ indicating direction dependence of the microwave absorption because the absorption coefficient $\alpha(\omega)$ is related to $N(\omega)$ as,
\begin{eqnarray*}
\alpha(\omega)=\frac{2\omega\kappa(\omega)}{c} 
\propto \omega {\rm Im}N(\omega),
\end{eqnarray*}
and the absorption intensity is given by $I(\omega)$=$I_0 \exp[-\alpha(\omega) l]$ where $l$ is the sample thickness. The nonreciprocal absorption $\Delta\alpha(\omega)=\alpha_{+}(\omega) - \alpha_{-}(\omega)$ represents the magnitude of NDD where $\alpha_{+}$ and $\alpha_{-}$ are absorption coefficients for microwaves propagating in the positive and negative directions, respectively.

In order to evaluate $N(\omega)$ and $\alpha_{\pm}(\omega)$ quantitatively, we need to calculate not only the dielectric and magnetic susceptibilities but also the following dynamical ME susceptibilities,
\begin{eqnarray}
\chi^{\rm em}_{\alpha \beta}(\omega) =
\frac{P_{\alpha}^{\omega}}{\sqrt{\epsilon_0 \mu_0}H_{\beta}^{\omega}},
\;\;\;
\chi^{\rm me}_{\alpha \beta}(\omega) =
\sqrt{\frac{\mu_0}{\epsilon_0}}
\frac{M_{\alpha}^{\omega}}{E_{\beta}^{\omega}}.
\end{eqnarray}
For the values of $\epsilon_{zz}^{\infty}$ and $\epsilon_{yy}^{\infty}$ in Eqs.~(\ref{eqn:CRI1}) and (\ref{eqn:CRI2}), we assume an isotropic dielectric tensor, i.e, $\epsilon_{zz}^{\infty}$=$\epsilon_{yy}^{\infty}$=$\epsilon^{\infty}$ for simplicity, and set $\epsilon^{\infty}$=8 according to the dielectric-measurement data~\cite{Belesi12,Miller10}. In turn, we take $\mu_{zz}^{\infty}$=$\mu_{yy}^{\infty}$=1 for permeability. The value of $J$ is set to be $J$=1 meV so as to reproduce the experimental $T_{\rm c}$ for the SkX-paramagnetic transition.

In Fig.~\ref{Fig4}(a) [Fig.~\ref{Fig4}(b)], we display calculated $\omega$ dependence of the absorption coefficients, $\alpha_+(\omega)$ and $\alpha_-(\omega)$, at several values of $H_z$ for $\bm E^\omega$$\parallel$$\bm z$ and $\bm H^\omega$$\parallel$$\bm y$ [$\bm E^\omega$$\parallel$$\bm y$ and $\bm H^\omega$$\parallel$$\bm z$]. The out-of-plane $\bm E^\omega$$\parallel$$\bm z$ and in-plane $\bm H^\omega$$\parallel$$\bm y$ activate two rotation modes with opposite senses~\cite{Mochizuki12,Petrova11}, i.e., lower-lying counterclockwise and higher-lying clockwise modes, which give two spectral peaks in Fig.~\ref{Fig4}(a). 
We find that for the lower-lying resonance, the $\Delta\alpha(\omega)$ increases as $H_z$ increases, and reaches more than 0.25 cm$^{-1}$, which corresponds to a relative change $\Delta\alpha$/$\alpha_{\rm ave}$=2($\alpha_+ - \alpha_-$)/($\alpha_+ + \alpha_-$)$\sim$20 $\%$ at maximum. On the other hand, both the in-plane $\bm E^\omega$$\parallel$$\bm y$ and out-of-plane $\bm H^\omega$$\parallel$$\bm z$ activate a breathing mode~\cite{Mochizuki12,Petrova11}, which give a single spetral peak in Fig.~\ref{Fig4}(b). Again the $\Delta\alpha(\omega)$ increases with increasing $H_z$, and reaches approximately 0.14 cm$^{-1}$, corresponding to $\Delta\alpha$/$\alpha_{\rm ave}$ of 10 $\%$. These values do not depend on the value of $\alpha_{\rm G}$. Such a huge directional dichroism is quite rare for any frequency range~\cite{Kezsmarki11,Takahashi12,Bordacs12,Miyahara11}, and has never been realized at GHz frequencies. This is because most of the multiferroics with simple magnetic orders have resonant frequencies much larger than microwave frequencies due to the large spin gaps. In turn, the long-period skyrmion textures have small spin gaps and their nontrivial collective modes with GHz frequencies enable us to achieve interesting microwave functions.

In summary, we have theoretically predicted that the SkX phase of chiral-lattice insulator Cu$_2$OSeO$_3$ shows the enhanced diode effect on linearly polarized microwaves as a consequence of interference between magnetic and electric responses from the multiferroic skyrmion texture.
Our prediction demonstrates that the skyrmion and the SkX host interesting dynamical phenomena in the microwave frequency regimes in addition to the peculiar transport and spintronics phenomena.

The authors thank A. Rosch, N. Nagaosa, Y. Tokura, and N. Furukawa for discussions. M.M. thanks the University of Cologne for hospitality. This work was supported by the Japan Society for the Promotion of Science (JSPS) through the `Funding Program for World-Leading Innovative R\&D on Science and Technology (FIRST Program)', by G-COE Program ``Physical Sciences Frontier" from MEXT Japan, by PRESTO program of JST, and by Murata Science Foundation.


\end{document}